\documentstyle[preprint,prl,aps]{revtex}
\begin{document}
\draft
\title{Spiral magnets as gapless Mott insulators}
\author{R. C\^ot\'e and A.-M.S. Tremblay}
\address{Centre de Recherche en Physique du Solide et D\'epartement de Physique}
\address{Universit\'e de Sherbrooke, Sherbrooke, Qu\'ebec, Canada J1K 2R1}
\date{\today }
\maketitle

\begin{abstract}
In the large $U$ limit, the ground state of the half-filled,
nearest-neighbor Hubbard model on the triangular lattice is the
three-sublattice antiferromagnet. In sharp contrast with the square-lattice
case, where transverse spin-waves and charge excitations remain decoupled to
all orders in $t/U$, it is shown that beyond leading order in $t/U$ the
three Goldstone modes on the triangular lattice are a linear combination of
spin and charge. This leads to non-vanishing conductivity at any finite
frequency, even though the magnet remains insulating at zero frequency. More
generally, non-collinear spin order should lead to such gapless insulating
behavior.
\end{abstract}



In band theory, insulating behavior is due to the existence of
an energy gap in the single-particle excitation spectrum. It is well known,
however, that interactions neglected in band theory may lead to
insulating properties, a point first discussed by Mott 
and, in 1964,
by Kohn who showed that the existence of an energy gap is sufficient
but not necessary to have an insulator
\cite{kohn}. 
An insulator is thus best defined by a vanishing
zero-temperature {\it DC} conductivity.
In the usual examples of Mott insulators, however,
there {\it is} generally
a gap not only in the single-particle excitations,
but also in 
the frequency dependent conductivity which is
related to excited states at constant number of particles (particle-hole
excited states instead of single-particle excitations). 
 In this letter, we show that when interactions lead to
magnetic order in {\it itinerant} spin $1/2$ systems, the resulting Mott
insulator generally has {\it no} gap in the frequency dependent
conductivity: This is the behavior of a gapless insulator. It is only when the
magnetic order is collinear antiferromagnetism on hypercubic lattices that
one can demonstrate the existence of a gap in the conductivity. In the
general case that we consider in detail, the mixed spin and charge character
\cite{note:John} of the finite ${\bf q}$ {\it collective }(Goldstone) modes
leads to the vanishing of the gap in the conductivity for a finite range of
interaction strengths, despite the fact that the {\it single-particle}
excitations do have a gap.

We start from the general one-band Hubbard model 
\begin{equation}
\label{un}H=-\sum_{\left\langle i,j\right\rangle ,\sigma }t_{ij}\left(
c_{i\sigma }^{\dagger }c_{j\sigma }+c_{j\sigma }^{\dagger }c_{i\sigma
}\right) +U\sum_in_{i\uparrow }n_{i\downarrow }\;, 
\end{equation}
where the first sum is over pairs of sites, $t_{ij}$ is the hopping integral
and $U$ the on-site Coulomb interaction. We then define 
\begin{equation}
\label{douzep}S^\mu (i)\equiv \frac 12\sum_{\alpha ,\beta }c_{i\alpha
}\sigma _{\alpha \beta }^\mu c_{i\beta }\;, 
\end{equation}
a dimensionless four-component spin and charge operator.(To be explicit, we
let the matrix index $\mu $ take the values $\rho ,x,y,z$, and we define $%
\sigma ^\rho $ as the unit matrix.) We consider the general class of
magnetically ordered states with planar spiral order 
\begin{equation}
\label{deux}\langle S_i^z\rangle +i\langle S_i^x\rangle =Se^{i{\bf Q}\cdot 
{\bf R_i}}\;, 
\end{equation}
where $S$ is the order parameter representing the average moment on each
site. Without loss of generality, we take the spiral in the $x-z$ plane.
Well known examples of such states include the collinear
antiferromagnet on the square
lattice, ${\bf Q}=\pi /a\ \widehat{x}\ +\quad \pi /a\ \widehat{z}$, and the
three-sublattice antiferromagnet occurring on the frustrated triangular
lattice, a $120^0$ spiral with ${\bf Q}={4\pi /3a}\;\hat x$. As always when
working with planar spiral order, the analysis is greatly simplified by
using a rotating (orthonormal) basis in which the quantization axis at every
site points in the same direction as the average spin density of Eq.(\ref
{deux}). In this rotating frame, the single-particle Green's function $%
G_{\alpha \beta }(i,j;\tau )\equiv -\langle Tc_{i\alpha }(\tau )c_{j\beta
}^{\dagger }(0)\rangle $ becomes, in matrix notation, 
\begin{equation}
\label{cinq}\widetilde{G}(i,j;i\omega _n)=T_i^{\dagger }\;G(i,j;i\omega
_n)\;T_j\;, 
\end{equation}
where the rotation matrix in spin space $T_i=e^{-i({\bf Q}\cdot {\bf R_i}%
)\sigma ^y/2}$ (with $\sigma ^y$ the usual Pauli matrix) depends on the site 
$i$. In this reference frame, the Green's function matrix $\widetilde{G}$
has the full underlying lattice periodicity, even for incommensurate spiral
order. In the Hartree-Fock approximation, we find 
\begin{equation}
\label{six}\widetilde{G}({\bf k},i\omega _n)={\frac{A_{+}({\bf k})/2}{%
i\omega _n+\mu -E_{+}({\bf k})}}+{\frac{A_{-}({\bf k})/2}{i\omega _n+\mu
-E_{-}({\bf k})}}\;, 
\end{equation}
where

$$
A_{\pm }({\bf k)}\equiv \left( 
\begin{array}{cc}
1\mp \frac \Delta {E({\bf k)}} & \pm i 
\frac{\eta ({\bf k})}{E({\bf k)}} \\ \mp i\frac{\eta ({\bf k})}{E({\bf k)}}
& 1\pm \frac \Delta {E({\bf k)}} 
\end{array}
\right) 
$$
with the definitions%
$$
\epsilon _0({\bf k})\equiv -{\frac 1N}\sum_{i,j}t_{ij}\cos {\bf k\cdot (R}_i-%
{\bf R}_j) 
$$
for the single-particle dispersion in the paramagnetic phase ($N$ is the
number of sites), and 
\begin{equation}
\label{quatre}
\begin{array}{cc}
\epsilon _{\pm }\equiv \epsilon _0({\bf k\pm Q}/2) & \eta ( 
{\bf k})\equiv \frac{\epsilon _{+}({\bf k})-\epsilon _{-}({\bf k})}2 \\ 
\epsilon ({\bf k})\equiv \frac{\epsilon _{+}({\bf k})+\epsilon _{-}({\bf k})}%
2+\frac U2 & \Delta \equiv US \\ 
E({\bf k})\equiv \sqrt{\eta ^2({\bf k})+\Delta ^2} & E_{\pm }({\bf k})\equiv
\epsilon ({\bf k})\pm E({\bf k}) 
\end{array}
. 
\end{equation}
The gap equation is given by the self-consistency requirement, 
\begin{equation}
\label{neuf}\frac U{2N}\sum_{{\bf k}}{\frac{[f(E_{-}({\bf k}))-f(E_{+}({\bf k%
}))]}{E({\bf k})}}=1\;, 
\end{equation}
where $f(x)$ is the Fermi function and the wave vector ${\bf k}$ spans the
entire paramagnetic Brillouin zone of the crystal. The chemical potential is
determined by number conservation,%
$$
\frac 1N\sum_{{\bf k}}[f(E_{-}({\bf k}))+f(E_{+}({\bf k}))]=1. 
$$

Proceeding to the collective excitations in the ordered state, we define the
matrix response function%
$$
{\chi }^{\;\mu \nu }(i,j;\tau )\equiv -\left\langle TS^\mu (i;\tau )S^\nu
(j;0)\right\rangle +\left\langle S^\mu (i)\right\rangle \left\langle S^\nu
(j)\right\rangle . 
$$
$\chi $ is obtained in the Generalized Random-Phase Approximation (GRPA) by
the usual summation of bubble and ladder diagrams. 
Since the GRPA is conservative, going beyond this approximation should
not change the qualitative aspects of ours results.
In the rotating frame
where $\chi ^{\mu \nu }(i,j;\tau )\rightarrow \widetilde{\chi }^{\mu \nu
}(i-j;\tau )\;$, the matrix GRPA equation\cite{details} takes the form 
\begin{equation}
\label{treize}{\widetilde{\chi }}({\bf q},i\Omega _n)={\widetilde{\chi }}%
^{\;0}({\bf q},i\Omega _n)+2U{\widetilde{\chi }}^{\;0}({\bf q},i\Omega
_n)\Gamma {\widetilde{\chi }}({\bf q},i\Omega _n)\;, 
\end{equation}
where $\Gamma $ is a diagonal matrix with $\Gamma ^{\rho \rho }\equiv 1,\
\Gamma ^{xx}\equiv \Gamma ^{yy}\equiv \Gamma ^{zz}\equiv -1$.

At $T=0K$, which we will consider from now on, the retarded zeroth-order
matrix susceptibility ${\widetilde{\chi }}^{\;0}({\bf q},\omega +i\delta )$
is given by, 
\begin{equation}
\label{quinze}{\widetilde{\chi }}^{\;0}({\bf q},\omega +i\delta )=\frac 1{8N}%
\sum_{{\bf k}}\left( 
\begin{array}{cccc}
\left( \frac{-\Delta ^2+EE^{\prime }-\eta \eta ^{\prime }}{EE^{\prime }}%
\right) \Lambda _{-} & i\Delta \left( \frac{\eta -\eta ^{\prime }}{%
EE^{\prime }}\right) \Lambda _{-} & \left( \frac{-\eta }E+\frac{\eta
^{\prime }}{E^{\prime }}\right) \Lambda _{+} & \Delta \left( 
\frac{-1}E+\frac 1{E^{\prime }}\right) \Lambda _{+} \\ -i\Delta \left( \frac{%
\eta -\eta ^{\prime }}{EE^{\prime }}\right) \Lambda _{-} & \left( \frac{%
\Delta ^2+EE^{\prime }+\eta \eta ^{\prime }}{EE^{\prime }}\right) \Lambda
_{-} & i\Delta \left( \frac 1E+\frac 1{E^{\prime }}\right) \Lambda _{+} & 
-i\left( 
\frac \eta E+\frac{\eta ^{\prime }}{E^{\prime }}\right) \Lambda _{+} \\ 
\left( \frac{-\eta }E+\frac{\eta ^{\prime }}{E^{\prime }}\right) \Lambda
_{+} & -i\Delta \left( \frac 1E+\frac 1{E^{\prime }}\right) \Lambda _{+} & 
\left( \frac{\Delta ^2+EE^{\prime }-\eta \eta ^{\prime }}{EE^{\prime }}%
\right) \Lambda _{-} & -\Delta \left( 
\frac{\eta +\eta ^{\prime }}{EE^{\prime }}\right) \Lambda _{-} \\ \Delta
\left( \frac{-1}E+\frac 1{E^{\prime }}\right) \Lambda _{+} & i\left( \frac 
\eta E+\frac{\eta ^{\prime }}{E^{\prime }}\right) \Lambda _{+} & -\Delta
\left( \frac{\eta +\eta ^{\prime }}{EE^{\prime }}\right) \Lambda _{-} & 
\left( \frac{-\Delta ^2+EE^{\prime }+\eta \eta ^{\prime }}{EE^{\prime }}%
\right) \Lambda _{-} 
\end{array}
\right) 
\end{equation}
where unprimed functions are to be evaluated at ${\bf k}+{\bf q}/2$ and the
primed functions at ${\bf k}-{\bf q}/2$, and where $\Lambda _{\pm }\equiv
\gamma _{-}\pm \gamma _{+}$ with 
\begin{equation}
\label{seize}\gamma _{\pm }={\frac 1{\omega +i\delta -(\epsilon -\epsilon
^{\prime })\pm (E+E^{\prime })}}\;. 
\end{equation}

The response function in the rotating frame $\widetilde{\chi }$ can be
related to the laboratory response $\chi $ but the expressions are lengthy 
\cite{details}. For the density response however, the relation simply is 
\begin{equation}
\label{vdeux}\chi ^{\rho \rho }({\bf q},{\bf q^{\prime }},\omega +i\delta )=%
\widetilde{\chi }^{\rho \rho }({\bf q},\omega +i\delta )\delta _{{\bf q},%
{\bf q^{\prime }}}\;. 
\end{equation}
The other response functions involve linear combinations of $\widetilde{\chi 
}$ at different wave-vectors. 
The poles of $\widetilde{\chi }$, which of course coincide with those of $%
\chi $, give the position of the collective modes.

For hypercubic lattices in arbitrary dimension larger than one, symmetry
arguments can be used to show that whatever the hopping matrix $t_{ij}$ and
the value of $U$, as long as there is long-range collinear antiferromagnetic
order, the response matrix $\widetilde{\chi }$ is block diagonal. There are
then two Goldstone modes with purely transverse $x-y$ character. For the
nearest-neighbor model at half-filling, it is well known\cite{swz} that in
the limit $t\ll U$, the above approach correctly reproduces the standard
spin wave result of the Heisenberg model with $J\sim 4t^2/U$. For
next-nearest neighbor hopping on the square lattice, the resulting spin
Hamiltonian in the large $U$ limit is frustrated. Nevertheless, it follows
from above that as long as there is long-range antiferromagnetic order,
there are two Goldstone modes with purely transverse $x-y$ character, as in
the simpler non-frustrated case.

From now on, we restrict ourselves to the nearest-neighbor triangular
lattice at half-filling. On this lattice, the frustrated antiferromagnetism
can give rise to non-collinear magnetic order, a subject of current interest
in itself\cite{plumer}. Much work has been done to obtain the magnetic phase
diagram as the electrons are allowed to be more and more itinerant ($i.e.$
as the ratio $U/t$ decreases)\cite{ikawa}.
Krishnamurthy $et\;al.$\cite{krishna} and Jayaprakash $et\;al.$\cite{jaya}
have shown that, at half-filling, as $U$ is increased the Hartree-Fock
ground-state evolves from a paramagnetic metal to a metallic incommensurate
spiral Spin-Density-Wave (SDW) then to a commensurate linear SDW with
indirect gap and finally, at higher values of $U$, into a spiral SDW
insulator with a three-sublattice $120^{\circ }$ twist between spins on
neighboring sites. This spiral (or equivalently helical) SDW state is the
well known ground state of the Heisenberg model (with $J\sim 4t^2/U$). We
call this state the $120^{\circ }$ spiral SDW phase.

We go beyond previous studies by accounting for the collective excitations
of the $120^{\circ }$ spiral SDW phase. The electronic dispersion in the
absence of interaction for the nearest-neighbor model on the triangular
lattice is given\cite{signe} by $\epsilon _0({\bf k})=-2t(\cos (k_xa)+2\cos (%
{k_xa/2})\cos (\sqrt{3}k_ya/2))$. In the paramagnetic Brillouin zone, the
single band $\epsilon _0({\bf k})$ is split into the two subbands $E_{\pm }(%
{\bf k})$ by the presence of spiral order, as can be seen from Eqs.(\ref{six}%
),(\ref{quatre}). The direct single-particle
energy gap is $2\Delta $ but the indirect
gap $E_{IG}={\rm Min}[E_{+}({\bf k})]-{\rm Max}[E_{-}({\bf k})]$ can
be lower than this value. When the indirect gap vanishes, a
transition to a metallic state occurs. This transition takes place at a
value of $t/U\approx 0.195$, above which the order parameter $S$ decreases
very rapidly. However, the collective modes destabilize the insulating $%
120^0 $ spiral SDW\ phase at $t/U\approx 0.146$, well before the above
metallic transition is reached. In the following, we restrict ourselves to
the regime at half-filling where the indirect gap is positive and only the
lowest energy band is filled.

Expanding $\widetilde{\chi }^0({\bf q},\omega )$ Eq.(\ref{quinze}) to second
order in $\omega /U,t/U$, and using the gap equation to eliminate $S$ in
favor of $U$, we find after tedious algebra that to leading order in $t/U$,
the poles $\omega({\bf q})$ of the response function Eq.(\ref{treize})
correctly reproduce the standard spin wave result of the Heisenberg model
with $J=4t^2/U$\cite{naga}. If ${\bf q}$ is restricted to the magnetic
Brillouin zone, the three spin-wave branches (Goldstone modes) are given by $%
\omega ({\bf q}),\omega ({\bf q}+{\bf Q}),\omega ({\bf q}-{\bf Q})$.

As the ratio $t/U$ increases and the electrons become more itinerant,
higher-order terms in $t/U$ cannot be neglected and the coupling between
charge and the three spin components becomes important. The matrix $%
\widetilde{\chi }({\bf q},\omega )$ is no-longer block diagonal and all
response functions then share the same poles, although with different
weights. Three of the poles are still Goldstone modes with a vanishing
frequency at ${\bf q=0}$ in the {\it magnetic} Brillouin zone. In Fig.~1 we
plot the imaginary part of the response functions $\widetilde{\chi }^{\rho
\rho }$ and $\widetilde{\chi }^{xx}$ for various values of ${\bf q}$. We
plot in Fig.2 the dispersion relation of the collective modes obtained by
following the peak of the response function as ${\bf q}$ is varied along
high-symmetry directions.

It is clear from Fig.~1 that the coupling between spin and charge is very
small so that the spin waves are only very slightly modified by the charge
fluctuations. The intensity of this coupling increases with $t/U$ but there
is only a narrow range of $t/U$ over which the $120^0$ spiral SDW is stable.
Indeed, at $t/U\approx 0.146$ one of the collective modes softens,
destabilizing the $120^0$ spiral SDW ground state at a smaller value than
the $t/U=0.16$ found in Ref.\cite{jaya} using only self-consistent
Hartree-Fock solutions.

Because the collective modes with coupled spin-wave and charge extend to 
{\it zero-frequency}, it is natural to ask whether the system could becomes
conducting
despite the existence of a gap in the single-particle
excitations. The appropriate definition of an insulator is that it has a
vanishing Drude weight.\cite{kohn,scalapi} This statement is equivalent to
saying that the coefficient of the delta function $\delta (\omega )$ in the
zero-temperature DC conductivity $\sigma _{DC}=\lim _{\omega \to 0}\lim
_{q\to 0}\sigma (\omega ,{\bf q})$ vanishes. We obtained this quantity from
the appropriate limit of 
\begin{equation}
\label{vun}\sigma (\omega ,{\bf q})=ie^2{\frac{(\omega +i\delta )}{q^2}}\chi
^{\rho \rho }(\omega +i\delta ,{\bf q},{\bf q})\;.
\end{equation}
In this expression, one has to take the part of $\chi ^{\rho \rho }(\omega
+i\delta ,{\bf q},{\bf {q})=}\widetilde{\chi }^{\rho \rho }(\omega +i\delta ,%
{\bf q)}$ which is irreducible with respect to the interaction.\cite{Mahan}
This eliminates the effect of screening. Expanding to second order in ${\bf q%
},\omega $, we found after lengthy algebra that the charge-response function
decreases faster than $q^2$ for $q\to 0$ so that the Goldstone-mode
contributing to $\sigma (\omega ,{\bf q})$ has effectively zero weight at $%
{\bf q}\to 0$. The same conclusion is reached by calculating the transverse
current-current response function in the GRPA. It then follows that the
system remains insulating even though there is no gap in the charged
collective excitations. This is not so surprising since at ${\bf q}=0$ the
Goldstone modes restore rotational invariance: hence they have the same
parity as the ground state so that
matrix elements of the current operator between the ${\bf q}=0$
modes and the ground state vanish.

At finite wave vector and frequency, however, the conductivity is finite
and, as shown in Fig.~3, the absorption described by the real part is
exactly at the Goldstone-mode position, as expected. We conjecture that the
insulator can be changed into a bad conductor by impurities. Indeed,
impurities transform the single delta function $\delta (\omega )$ appearing
in the {\it DC }conductivity of perfect metals into a Lorentzian-like
response. In our case, the breaking of crystal momentum conservation by
impurities would probably, by a similar mechanism, broaden the finite $q$
resonances to make the $q=0$ conductivity finite everywhere in the
single-particle gap, including zero frequency. The question is difficult
because of possible Anderson localization.

In conclusion, the triangular three-sublattice antiferromagnet described by
the large $U$ half-filled one-band Hubbard model provides an example of a
gapless Mott insulator. In this system, the Goldstone modes have mixed spin
and charge character, leading to the disappearance of the gap in the
conductivity, despite a vanishing {\it DC }conductivity and the existence of
a gap in the single-particle excitations. This phenomenon is generic for
itinerant magnets with spiral order. An exception is when the ordering
wave vector corresponds to collinear antiferromagnets. 

We are grateful to A. Caill\'e, M. Plumer, D. S\'en\'echal and D. Boies for
useful discussions. This work was supported by the Natural Sciences and
Engineering Research Council of Canada, (NSERC) the Fonds pour la formation
de chercheurs et l'aide \`a la recherche from the Government of Qu\'ebec
(FCAR), and (A.-M.S.T.) the Canadian Institute of Advanced Research (CIAR).

%

\begin{figure}
\caption{ 
Imaginary part of the density and (inset) spin 
response functions at $t/U=0.1$. From left to right
the peaks, broadened by a small imaginary part, correspond to
$q_y=0,q_x=0.1,0.15,0.25$ (in units of $2\pi/a$)
} 
\end{figure}

\begin{figure}
\caption{ 
Dispersion relation of the mixed spin and charge collective
modes along the edges of the irreducible
Brillouin zone of the triangular lattice. The full, dashed
and dot-dashed curves are respectively for $t/U=0.1,0.12,0.146$. 
} 
\end{figure}

\begin{figure}
\caption{ 
Real part of the conductivity defined in Eq.~(13) at
$t/U=0.1$ and for the wave vectors 
$q_y=0;
 q_x=0.1,0.15,0.25$
(in units of $2\pi/a$) represented by the full, dashed and dot-dashed
curves respectively. 
} 
\end{figure}


\begin{references}
\bibitem{kohn}  W. Kohn, Phys. Rev. {\bf 133}, A171 (1964).

\bibitem{note:John}  Coupling between the transverse spin excitations and
charge excitations has been found in the {\it metallic }state of
incommensurate spiral SDW states on the square lattice by S. John, P.
Voruganti, and W. Goff, Phys. Rev. B {\bf 43}, 13 365 (1991).

\bibitem{details}  Details of the derivation will be given elsewhere.

\bibitem{swz}  J. R. Schrieffer, X. G. Wen, and S. C. Zhang, Phys. Rev. B 
{\bf 39}, 11663 (1989); A.V. Chubukov, and D. Frenkel, Phys. Rev. B 46,
11~884 (1992).

\bibitem{plumer}  
M.L. Plumer, A. Caill\'e, A. Mailhot and H.T. Diep in {\it Magnetic
Systems with Competing Interactions, }edited by H.T. Diep (World Scientific,
Singapore, 1994).

\bibitem{ikawa}  A. Ikawa and M. Ozaki, J. Phys.: Condens. Matter {\bf 4},
4039 (1992);
K. Machida, M. Fujita, Phys. Rev. B {\bf 42}, 2673
(1990); K. Machida, Surf. Science, {\bf 242}, 206 (1991);
M. Fujita, T. Nakanishi, K. Machida, Phys. Rev. B {\bf 45}%
, 2190 (1992);
C. Pinettes and C. Lacroix, Solid State Commun., {\bf 85}%
, 565 (1993).

\bibitem{krishna}  H. R. Krishnamurthy, C. Jayaprakash, S. Sarker, W.
Wenzel, Phys. Rev. Lett. {\bf 64}, 950 (1990).

\bibitem{jaya}  C. Jayaprakash, H. R. Krishnamurthu, S. Sarker, W. Wenzel,
Europhys. Lett. {\bf 15}, 265 (1991).

\bibitem{signe}  At half-filling, all the results that
we discuss do not depend on the sign of the hopping integral $t$.

\bibitem{naga}  T. Nagamiya, Solid State Phys. {\bf 20}, 305 (1967).

\bibitem{scalapi}  D. J. Scalapino, S. R. White and S. C. Zhang, Phys. Rev.
Lett. {\bf 68}, 2830 (1992).

\bibitem{Mahan}  G.D. Mahan,{\it \ Many-Particle Physics} (Plenum, New York,
1981), p. 208.
\end{references}
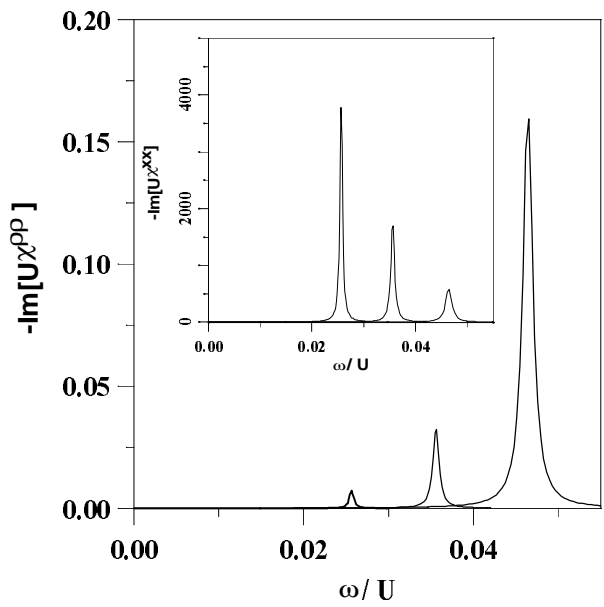
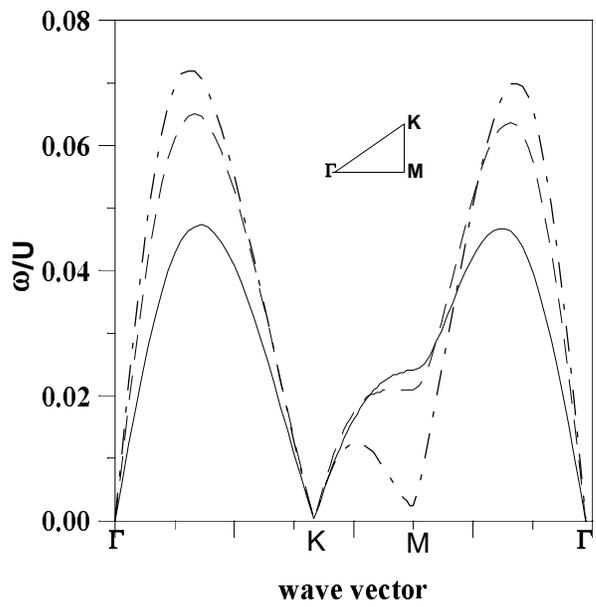
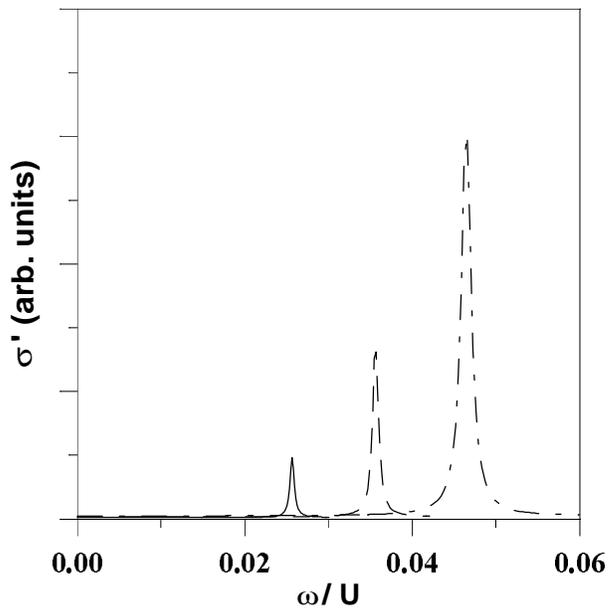
\end{document}